\documentclass[12pt]{iopart}
\usepackage{graphicx}
\usepackage{iopams}
\usepackage[pdfusetitle,
bookmarks=true,colorlinks=true, linkcolor=blue, citecolor=blue, linktoc=page]{hyperref}
\usepackage{comment}

\begin{document}
\title[]{Modelling of power exhaust in TCV positive and negative triangularity L-mode plasmas}

\author{E. Tonello$^{1,2}$, F. Mombelli$^{2}$, O. Février$^{1}$, G. Alberti$^{2}$, T. Bolzonella$^{3}$, G. Durr-Legoupil-Nicoud$^{1}$, S. Gorno$^{1}$, H. Reimerdes$^{1}$, C. Theiler$^{1}$, N. Vianello$^{3}$, M. Passoni$^{2,4}$, the TCV team$^{5}$ and the WPTE team$^{6}$}

\address{$^{1}$ Ecole Polytechnique Fédérale de Lausanne, Swiss Plasma Center, Lausanne, 1015, Switzerland}
\address{$^{2}$ Politecnico di Milano, Department of Energy, Milan, 20133, Italy}
\address{$^{3}$ Consorzio RFX, Padova, 35127, Italy}
\address{$^{4}$ Istituto per la Scienza e Tecnologia dei Plasmi, CNR, Milano, 20125, Italy}
\address{$^{5}$ See the author list of H. Reimerdes et al 2022 Nucl. Fusion 62 042018}
\address{$^{6}$ See the author list of E. Joffrin et al. Nucl. Fusion, 29th FEC Proceeding (2023)}

\ead{elena.tonello@epfl.ch}

\begin{abstract}
L-mode negative triangularity (NT) operation is a promising alternative to the positive triangularity (PT) H-mode as a high-confinement ELM-free operational regime. In this work, two TCV L-mode lower single null Ohmic discharges with opposite triangularity $\delta \simeq \pm 0.3$ are investigated using SOLPS-ITER modelling. The main focus is the exploration of the reasons behind the experimentally observed feature of NT plasmas being more difficult to detach than similar PT experiments~\cite{Fevrier2023}. SOLPS-ITER simulations are performed assuming the same anomalous diffusivity for particles $D_n^{AN}$ and energy $\kappa_{e/i}^{AN}$ in PT and NT. Nonetheless, the results clearly show dissimilar transport and accumulation of neutral particles in the scrape-off layer (SOL) of the two configurations, which consequently gives rise to different ionization sources for the plasma and produces different poloidal and cross-field fluxes. Simulations also recover the experimental feature of the outer target being hotter in the NT scenario (with $T_{e, NT} \gtrsim 5 \, \mathrm{eV}$) than in the PT counterpart. 
\end{abstract}

\section{Introduction}
\label{sec:intro}
Over the last decades, the so-called high-confinement mode (H-mode)~\cite{Wagner1982, ASDEX1989} was the regime of choice to fulfil confinement constrain compatible with nuclear fusion reactor operation. The design of ITER baseline scenario~\cite{Sips2018, ITER_RP2018} relies on H-mode operation. This regime is characterised by high core plasma pressure led by the formation of a transport barrier in the plasma edge. The presence of steep gradients across the separatrix is generally accompanied by the onset of Edge Localised Modes (ELMs)~\cite{Leonard2014}, periodic magneto-hydrodynamic (MHD) instabilities resulting in the ejection of energy and particles into the scrape-off layer (SOL). These result in particle and power loads onto the divertor, which concern ITER and steady-state reactor operation~\cite{eich_elm_2017}. For this reason, the quest towards the commercialisation of magnetic confinement nuclear fusion is eager to identify H-mode grade confinement configurations with small or no-ELMs~\cite{viezzer_prospects_2023}. One of the possibilities proven to be effective towards this goal is the exploitation of the effects of plasma shaping and, in particular, triangularity - denoted by $\delta$. The possibility of operating a tokamak reactor with a negative triangularity (NT) plasma shape has gained a lot of interest in recent years. Experiments performed in TCV~\cite{Pochelon1999, Camenen2007, coda_enhanced_2022}, DIII-D~\cite{ marinoni_h-mode_2019, Austin2019,  marinoni_diverted_2021} and ASDEX Upgrade~\cite{happel_overview_2023} tokamaks indicated that NT plasmas may offer a confinement level comparable to that of usual positive triangularity (PT) H-mode while working in an inherently ELM-free L-mode regime \cite{coda_enhanced_2022, fontana_effects_2020, austin_achievement_2019}. Therefore, NT operation might represent a viable and promising solution for future reactors in terms of confinement~\cite{ Kikuchi2019}.  \\
If negative triangularity is to be considered for reactor operation, a careful investigation of power exhaust and plasma detachment is crucial~\cite{krasheninnikov_physics_2017}. The latter was the goal of recent experiments performed in TCV, where L-mode scenarios with different triangularity and divertor geometry were compared to investigate the role of triangularity in detachment development~\cite{Fevrier2023}. The outcomes of these experiments indicate that the outer divertor target tends to remain hotter in NT compared to PT throughout L-mode density ramp experiments. This indicates that detachment is more difficult in NT compared to PT. The main aim of the work presented in this paper is to analyse the different power exhaust behaviour observed experimentally between PT and NT scenarios in light of the results of SOLPS-ITER simulations~\cite{wiesen_new_2015, BONNIN2016}. \\
The role of triangularity on energy confinement has been investigated by local~\cite{Merlo2015, Merlo2019} and global gyrokinetic simulations~\cite{Giannatale2022} and, in limited plasma shapes, also with turbulent fluid codes~\cite{laribi_impact_2021, Riva2017, Riva2020}. Power exhaust and edge plasma modelling diverted configurations, on the contrary, are still novel and largely unexplored fields in the context of comparing PT and NT plasma scenarios~\cite{muscente_analysis_2023}. The present work continues in the direction of investigating the effect of triangularities on the tokamak power exhaust issue, presenting a novel numerical study which compares simulations of TCV L-mode plasmas with opposite triangularity. To complement what has been presented in~\cite{muscente_analysis_2023}, this work investigates the role of triangularity on the transport and dissipation of particles, momentum and heat in the scrape-off layer (SOL) and towards the divertor targets by modelling NT and PT scenarios at fixed anomalous particle and heat diffusion, i.e. avoiding any assumptions on the different transport behaviour in the two scenarios. This allowed us to identify and isolate some of the experimentally observed behaviours that do not depend on turbulent transport across the separatrix but merely on dissimilarities related to differently shaped magnetic field lines. 

The paper is organised as follows: Section~\ref{sec:SOLPSsetup} presents SOLPS-ITER and highlights the setup parameters used in this work; Section~\ref{sec:results} presents and discusses the main results of this work focusing on the validation of SOLPS-ITER simulations against experimental data for low density attached conditions (Section~\ref{subsec:RefSim}). Discrepancies between the simulation results and experimental data are studied highlighting possible effects of the simplifying hypotheses used in the modelling, like inner-outer target asymmetries arising from drifts (Section~\ref{subsec:RevField}) or 3D geometrical effects (Section~\ref{subsec:3Deffects}). Section~\ref{subsec:diffPTNT} addresses the differences between the NT and PT reference simulations. Finally, Section~\ref{subsec:densityRamp} presents the results of a density ramp analysis; Section~\ref{sec:conclusions} draws the main conclusions and presents perspectives for future work. 

\section{Simulation setup}
\label{sec:SOLPSsetup}

\begin{figure*}
\centering
\includegraphics[width=\textwidth]{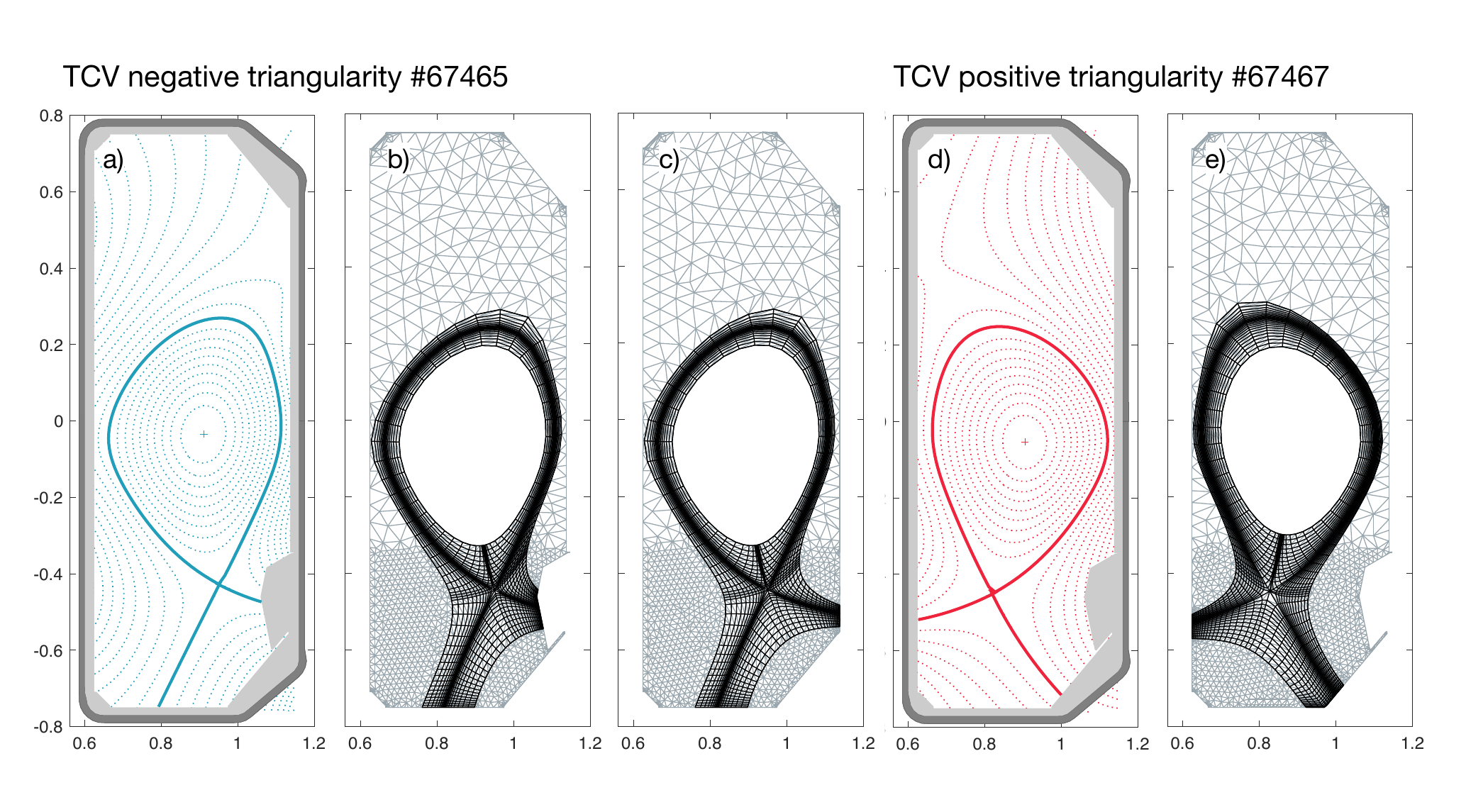}
\caption{ Reconstruction of the magnetic equilibria and computational meshes. B2.5 mesh in black and EIRENE mesh in grey. \textbf{a)} Negative triangularity equilibrium; \textbf{b)} NT computational mesh with port protection tiles and \textbf{c)} without port protection tiles. \textbf{d)} Positive triangularity equilibrium; \textbf{e)} PT computational mesh.} 
\label{fig:geos} 
\end{figure*}

This work makes use of the boundary plasma code suite SOLPS-ITER v. 3.0.8. This software package merges two main modules: B2.5, a 2D multi-fluid, mean-field plasma transport code, and a 3D kinetic Monte Carlo neutral transport code. The overall simulation domain is a 2D poloidal cross-section of the tokamak chamber that resembles the vacuum vessel shape. In this version of the code, plasma transport equations are solved on a structured quadrangular plasma mesh based on the magnetic equilibrium reconstruction. Charged species can only extend radially from the confined edge region, $\rho_\mathrm{pol} \simeq 0.9$, to the outer flux surface tangent of a non-divertor first wall structure. \\
The computational domains used for the simulations presented in this work are shown in figure~\ref{fig:geos}, together with the magnetic equilibrium reconstructions on which they are based. EIRENE triangular mesh, i.e. the neutral particle domain, is displayed in grey and the B2.5 plasma mesh in black. NT equilibrium and computational domain are shown in figures~\ref{fig:geos}.a), b) and c), while the PT ones are reported in figures~\ref{fig:geos}.d) and e). For both PT and NT, the B2.5 mesh resolution is $n_x \times n_y = 84 \times 24$ where $n_x$ and $n_y$ are the poloidal and radial resolutions, respectively. \\
The same boundary conditions for both NT and PT simulations were used to solve B2.5 equations:
\begin{itemize}
\item \textit{Divertor targets}: sheath boundary conditions, for densities, temperatures, velocities and electrostatic potential.
\item \textit{Far SOL}: zero gradients of the parallel velocities, zero current for the potential equation and decay boundary condition for $n_a$, $T_e$ and $T_i$ with $\lambda_\mathrm{decay} = 1 \, \mathrm{cm}$.
\item \textit{PFR}: zero gradients of the parallel velocities, zero current for the potential equation and leakage boundary condition for $n_a$, $T_e$ and $T_i$, e.g. for density $\Gamma_\mathrm{D^+, PFR} = \alpha_\mathrm{leak} n_\mathrm{D^+} c_S$ with $\alpha_\mathrm{leak} = - 1\times 10^{-3}$ and $c_S$ the ion sound velocity. 
\item \textit{Core}: at the core boundary, we used zero parallel velocity gradient and zero current boundary conditions. To solve density continuity equations we imposed a null particle flux crossing the core boundary $\Gamma_\mathrm{core, D^{+}} = -( \Gamma_\mathrm{core, D_{2}} + \Gamma_\mathrm{core,  D} +  \Gamma_\mathrm{core, D_{2}^+})$. The electron density at the core boundary $n_{e,\mathrm{core}}$ depends thus on the throughput strength (balance of puffing and pumping) and the value of ion radial diffusion $D_{n}^{AN}$. For electron and ion energy equations we set the power entering the edge from the core assuming $P_\mathrm{e, core} = P_\mathrm{i, core} \simeq 1/2 \times (P_\mathrm{Ohm} - P_\mathrm{rad, core}) = 200$ kW. 
\end{itemize}
SOLPS-ITER applies a mean-field edge plasma model which describes the cross-field anomalous transport as an average diffusion process. Anomalous diffusion of particles and electron and ion energies is characterised by the diffusion coefficients $D_{n}^{AN}$, $\kappa_{e}^{AN}$ and $\kappa_{i}^{AN}$, respectively. In this work, all the cross-field coefficients were uniform in space with $D_{n}^{AN} = 0.2 \, \mathrm{m^2 s^{-1}}$ and  $\kappa_{e}^{AN} = \kappa_{i}^{AN} = 1 \, \mathrm{m^2 s^{-1}}$, consistently with previous works on SOLPS modelling of TCV~\cite{Wensing2019, Wensing2021}. In particular, if not otherwise specified, these coefficients were kept fixed for both NT and PT scenarios. \\
To control the puffing of $D_2$ molecules a feedback scheme was applied. The value of $\Gamma_\mathrm{D_2}$ is internally varied by the code to make the plasma state converge to the desired value of electron density at the outer midplane (OMP) separatrix $n_{e, \mathrm{sep}}$. The reference PT and NT simulations are done with $n_{e, \mathrm{sep}} = 1.0 \times 10^{19} \, \mathrm{m}^{-3}$, corresponding to attached divertor conditions. \\
Carbon (C) impurities eroded from the wall by chemical and physical sputtering were first neglected to reduce the computational cost of the simulations. Simulations with carbon will be presented in section~\ref{subsec:densityRamp} to evaluate the impact of C impurities on detachment development. In the pure deuterium (D-only) simulations, we considered $D^+$, $D$, $D_2$, $D_2^{+}$. When carbon impurities are included (D+C), also neutral $C$ atoms and the six ionization states $C^{n+}$ were included. As far as ion temperature is considered, however, SOLPS-ITER uses the hypothesis of thermalised ions, i.e. a single ion temperature $T_i$ equation is solved. \\
Fluxes related to $\mathrm{E \times B}$, diamagnetic and viscous drifts, together with the corresponding currents, were also neglected in this work (no-drifts approximation). The main effect of drift velocities in the SOL is to produce asymmetries in transport between the inner and outer divertor. The no-drift approximation, thus, prevented us from capturing these features with our simulations. 

\section{Validation of SOLPS-ITER simulations against PT and NT TCV data}
\label{sec:results}
As discussed in section~\ref{sec:intro}, the main aim of the simulations presented in this work is to support the interpretation of the recent experiments performed in TCV which highlighted a different detachment behaviour between positive and negative triangularity L-mode discharges~\cite{Fevrier2023}. \\
The strategy adopted to perform SOLPS-ITER modelling aimed at investigating the \textit{geometrical} role of triangularity to power exhaust. Since SOLPS-ITER does not consistently include the physics behind cross-field turbulent transport, we avoided ad-hoc assumptions on differences between PT and NT cross-field diffusion coefficients and we used the same $D_n^{AN}$, $\kappa_e^{AN}$ and $\kappa_i^{AN}$ for both configurations. Simulations were performed using two different magnetic equilibria,  figure~\ref{fig:geos}.a and~\ref{fig:geos}.d, to build the plasma meshes, figures~\ref{fig:geos}.b and~\ref{fig:geos}.e. Besides anomalous diffusivities, both NT and PT simulations also employed the same input parameters in terms of boundary conditions and input power. 

\subsection{Reference simulations in attached conditions}
\label{subsec:RefSim}
\begin{figure*}
\centering
\includegraphics[width=0.75\textwidth]{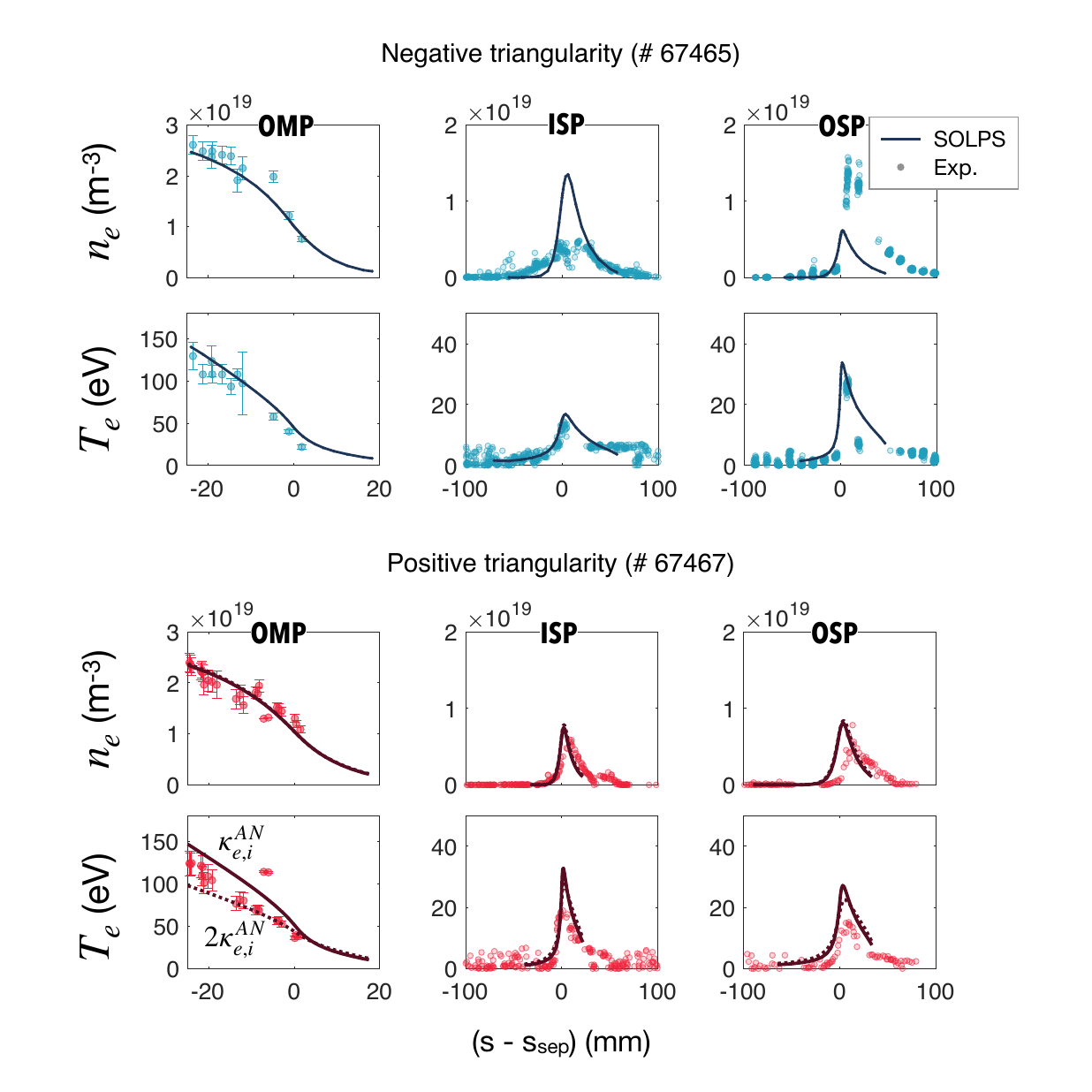}
\caption{Comparison between SOLPS-ITER simulations (solid lines) and TCV experimental data (points) for reference conditions. Electron density and temperature profiles for the negative (top-blue) and positive triangularity (bottom-red) discharges are shown at the outer midplane (OMP), inner strike-point (ISP) and outer strike-point (OSP). For PT, the dotted line represents the result of a simulation with electron and ion thermal diffusivity twice the reference value. } 
\label{fig:NT_PTref} 
\end{figure*}

We started our analysis by optimising the SOLPS input parameters for the negative triangularity configuration. 
The TCV discharge used as a reference for NT ($\# 67465$) is an Ohmic discharge with a lower-single null (LSN) configuration and reversed magnetic field $B_\varphi = 1.4 \, \mathrm{T}$. The reversed $B_\varphi$ direction, also called \textit{unfavourable} $\nabla B$, is the one for which the $\nabla B$ drift points upward and it allows to operate at high density and high current without transitioning to Ohmic H-mode~\cite{Fevrier2020}. The plasma current is $I_p = 220 \, \mathrm{kA}$ and upper and lower triangularities are fixed around $\delta \simeq - 0.27$. The experimental data that were compared with simulations are the Thompson scattering (TS) measurements, to compare the upstream (OMP) profiles and the wall-embedded Langmuir probes (LPs) to compare target profiles~\cite{Fvrier2021,DeOliveira2022}. \\
After a preliminary parametric scan of input power, anomalous diffusivities and boundary conditions, the input setup described in section~\ref{sec:SOLPSsetup} was chosen to match experimental data at the OMP and targets. 
The result of this comparison is shown in the top part of figure~\ref{fig:NT_PTref}. We observe a good agreement between the simulated profiles at the OMP and the TS data for electron temperature and density. The electron temperature profiles also agree well with LP data both at the inner (ISP) and outer strike points (OSP). As far as the electron density profiles are concerned, the differences observed between experimental data and simulations at the strike points are better addressed in section~\ref{subsec:RevField}.\\
The reference PT experiment ($\# 67467$) is very similar to the NT one regarding global discharge parameters. It is an Ohmic LSN discharge with reverse field $B_\varphi = 1.4$ T and plasma current $I_p = 220 \, \mathrm{kA}$. The triangularities (upper and lower) are positive with an absolute value similar to the NT case $\delta \simeq 0.27$.  Analogously to what was done for NT, we produced the reference simulation for the PT scenario. For this simulation, however, we did not perform any input optimisation and we kept fixed all the setup parameters used for the NT case. The only difference with the NT reference simulation is the plasma geometry and the computational mesh built on the PT equilibrium.
The comparison between PT simulation and experiment is shown in the bottom part of figure~\ref{fig:NT_PTref}. Even without assuming different anomalous diffusion coefficients compared to NT, a reasonable qualitative and quantitative agreement is obtained between simulation and experiment. In particular, the simulation matches within the experimental error the density profiles both at the OMP and at the strike points. SOLPS results, however, overestimate the experimental $T_e$, both upstream and at the targets. Assuming higher anomalous electron heat diffusivity $\kappa^{AN}_{e}$ for the PT case, agreement between simulations and experiments improves. The dotted line in figure~\ref{fig:NT_PTref} shows the result of a simulation performed with $\kappa^{AN}_{e}  = \kappa^{AN}_{i} = 2 \, \mathrm{m^2 s^{-1}}$, i.e. twice the reference value (solid line). This result agrees with the previous finding of reduced electron heat transport with more negative triangularity~\cite{Camenen2007}. 

\subsubsection{Forward and reversed $\vec B_\varphi$.}
\label{subsec:RevField}

\begin{figure*}
\centering
\includegraphics[width=1\textwidth]{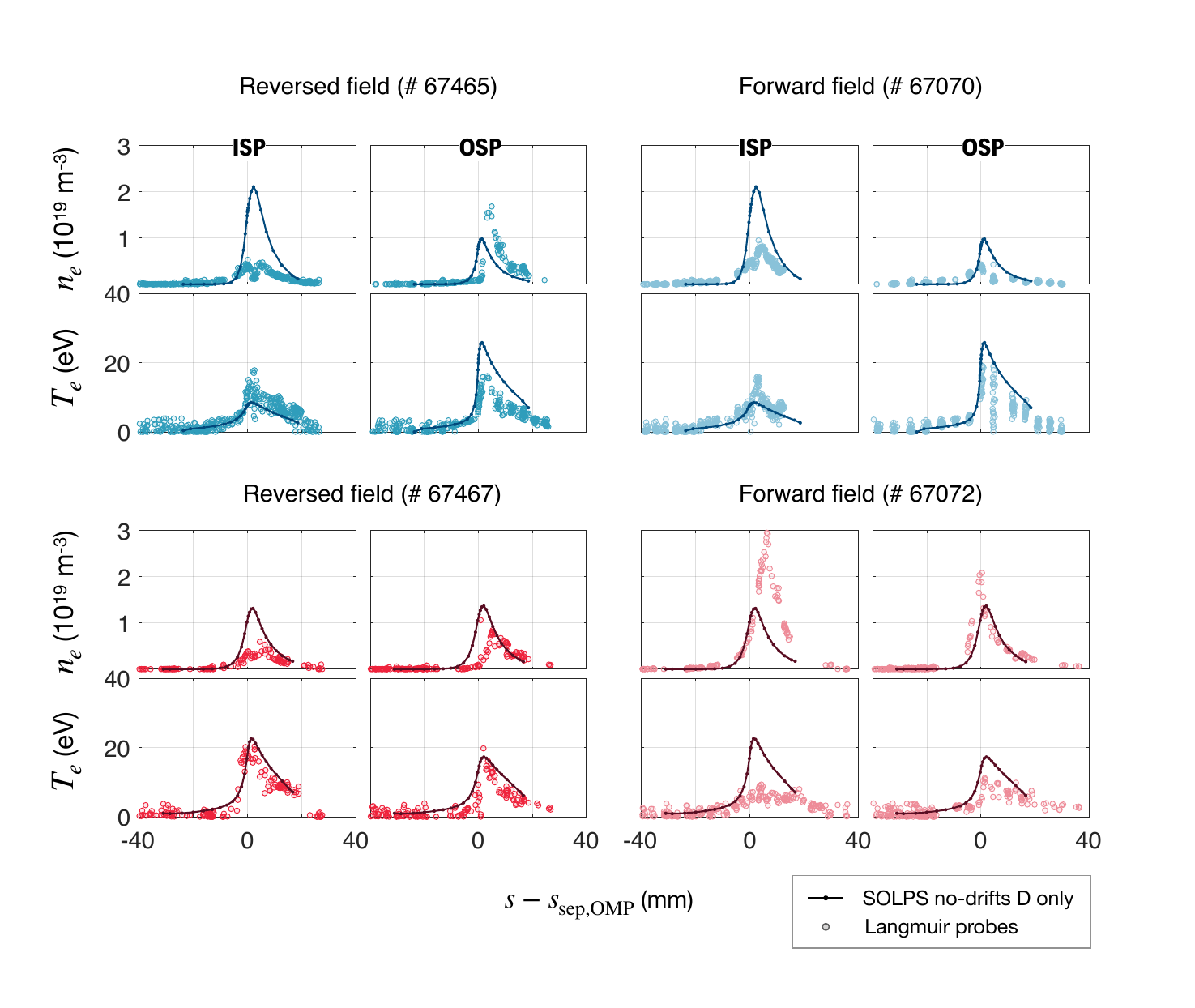}
\caption{Strike points electron density and temperature profiles in NT (top) and PT (bottom) discharges with reversed (left) and forward (right) $B_\varphi$ field discharges. Simulation and experimental data are taken at $n_{e, \mathrm{sep}} = 1.25 \times 10^{19} \, \mathrm{m^{-3}}$. The SOLPS-ITER results without drifts - hence identical for both reversed and forward $B_\varphi$ - are shown with solid lines while LP data are shown with dots.} 
\label{fig:revField} 
\end{figure*}

In the following two sections, we investigate the discrepancies between SOLPS-ITER reference simulations presented in section~\ref{subsec:RefSim} in light of the simplifying hypotheses used in our simulations. This section starts by highlighting how the absence of drift velocities in our simulations may affect our capabilities of reproducing target plasma profiles, simultaneously at the inner and outer targets. To study this, we compare the target profiles of the NT and PT reference simulations - without drifts - with LP data of TCV discharges with opposite $B_\varphi$ directions. Results of this comparison are shown in figure~\ref{fig:revField} for NT in the top part of the figure and for PT in the bottom one. For each of the four possible combinations of triangularity and $B_\varphi$ direction, we compared SOLPS results with the electron density and temperature profiles retrieved by LP measurements close to the inner and outer strike points. \\
The reversed field (RF) LP data come from the reference shots described in section~\ref{subsec:RefSim} - shot $\# 67465$ and $\# 67467$ for NT and PT, respectively. As forward field (FF) reference we used shots $\# 67070$ for NT and $\# 67072$ for PT. In these shots, the plasma current is again $I_p = 220 \, \mathrm{kA}$, triangularity  $\delta \simeq \mp 0.28$, respectively, and $B_\varphi = - 1.4$ T has the same magnitude but opposite direction to the reference ones. Simulations and experiments are compared at $n_{e, \mathrm{sep}}= 1.25 \times 10^{19} \, \mathrm{m^{-3}}$, instead of the reference $n_{e, \mathrm{sep}}$ value used in section~\ref{subsec:RefSim} because of the lack of lower density data for the NT-FF shot. \\
By comparing SOLPS results with RF and FF experimental data, we can first observe that the density profiles are generally more affected by the drift direction than the temperature profiles. Modifications in the latter, indeed, seem to be a direct consequence of the modification in the plasma density in front of the target and thus of the recycled flux. In particular, looking at the comparison between RF and FF cases in PT (bottom part of figure~\ref{fig:revField}), we observe that the field reversal from RF to FF increases the electron density in front of the inner target, thus increasing the recycling and reducing the electron temperature. In addition, in PT the field reversal has a small effect on the outer target profiles. This is compatible with previous findings of TCV simulations including drifts~\cite{Christen2017}, which highlighted the role of the $E \times B$ drift in increasing the plasma density in front of the inner target. \\
In NT, it is interesting to notice that changing the direction of $B_\varphi$ from RF to FF seems to have a different - and for some aspects opposite - effect compared to what is observed for PT. Indeed, the top part of figure~\ref{fig:revField} shows that in NT the $B_\varphi$ direction impacts mostly the density profiles at the OSP. Moreover, opposite to what happens at the ISP for PT, moving from RF to FF the ion flux reaching the OSP decreases. 
A more detailed analysis of these effects including drifts is planned for future studies.\\
Since the present work presents simulations without drifts, most of the conclusions will be drawn considering the OSP $T_e$ profiles, which for both PT and NT is only marginally affected by the $B_\varphi$ orientation.

\subsubsection{Toroidal symmetry and the port protection tiles}
\label{subsec:3Deffects}
\begin{figure*}
\centering
\includegraphics[width=1\textwidth]{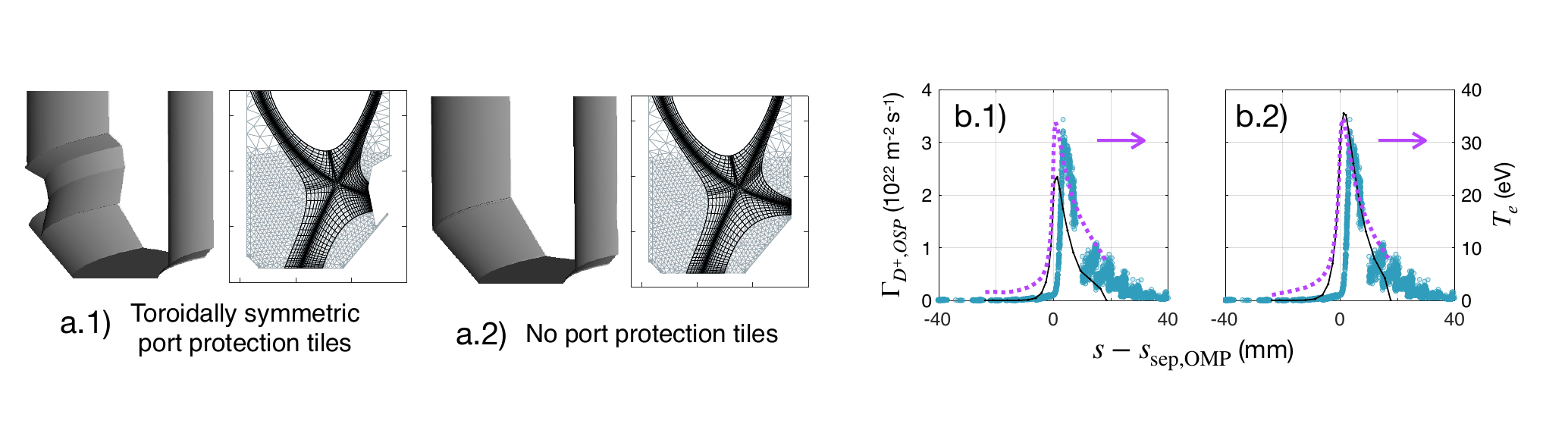}
\caption{Effect of the port protection tiles geometry on outer strike-point temperature profiles. a.1) Toroidally symmetric port-protection tile and b.1) the corresponding $\Gamma_{D^+}$ profile at the OSP; a.2) Geometry without port protection tile and b.2) the corresponding $\Gamma_{D^+}$ profile at the OSP. In figures b.1) and b.2) the solid black line represents the poloidal ion flux at the OSP, to be compared with the underlying LP data (dots), while the dotted violet line is the OSP $T_e$ profile.} 
\label{fig:PortGeo} 
\end{figure*}

Within a major upgrade of the first wall structure of TCV~\cite{Reimerdes2021}, in the lower-field side (LFS) region port protection tiles were added to shadow the lower lateral ports from direct plasma exposure. These tiles modify the cross-section of the first wall on the LFS in a non-toroidally symmetric way, as can be observed in figure 1.a of~\cite{Reimerdes2021}. Accounting for the correct shape of the port tiles may be particularly relevant for the NT scenario, where the OSP lies on top of these structures. However, since SOLPS-ITER assumes toroidal symmetry, we could only indirectly investigate their effect. To this aim, we performed two simulations: (i) one with the geometry shown in figure~\ref{fig:PortGeo}.a.1, with toroidally symmetric tiles (this is the reference NT simulation presented in section~\ref{subsec:RefSim}), and (ii) one with the geometry shown in figure~\ref{fig:PortGeo}.a.2 which neglects the presence of these structures and the OSP is located on the flat lateral surface of TCV rectangular vessel. The two full computational meshes used to perform these simulations are shown in figure~\ref{fig:geos}.b and~\ref{fig:geos}.c, respectively, they have the same spatial resolution and both simulations are performed using the input parameter discussed for the reference case in Section~\ref{sec:SOLPSsetup}. \\
Comparing SOLPS-ITER results, we observe that the upstream plasma profiles are unaffected by the change in the target geometry. At the outer target, the ion saturation current (solid line) is higher in the simulation without port protection tiles (figure~\ref{fig:PortGeo}.b.2) but the peak electron temperature  $\max{(T_{e, OSP})}$ (dotted line) is the same for both geometries. The different $\Gamma_{\mathrm{pol},D^+} \sim  n_{e, OSP} \sqrt{(T_{e, OSP}+T_{i, OSP})/m_i} B_\theta/B$ is due to a different $n_e$ in front of the outer target. The higher $\Gamma_{\mathrm{pol},D^+}$ in ~\ref{fig:PortGeo}.a.2 is explained by the different poloidal flux expansion $f_x \simeq R_{OMP} B_{\theta, OMP}/R_{OSP} B_{\theta, OSP}$~\cite{Theiler2017} of the two geometries. Indeed, $f_\mathrm{pol}$ in geometry~\ref{fig:PortGeo}.a.1 is approximately twice $f_\mathrm{pol}$ of~\ref{fig:PortGeo}.a.2. \\
In future studies, to have better insight and understanding of the differences between PT and NT concerning also ion fluxes and plasma density profiles at the outer target, e.g. in light of studies including the effect of drifts, it would be better to compare magnetic configurations with similar divertor geometries where the OSP is located on the flat bottom of TCV vessel also for the NT scenario. Such geometries have already been explored experimentally~\cite{Fevrier2023}.

\subsection{Differences between NT and PT reference scenarios}
\label{subsec:diffPTNT}
\begin{figure*}
\centering
\includegraphics[width=1\textwidth]{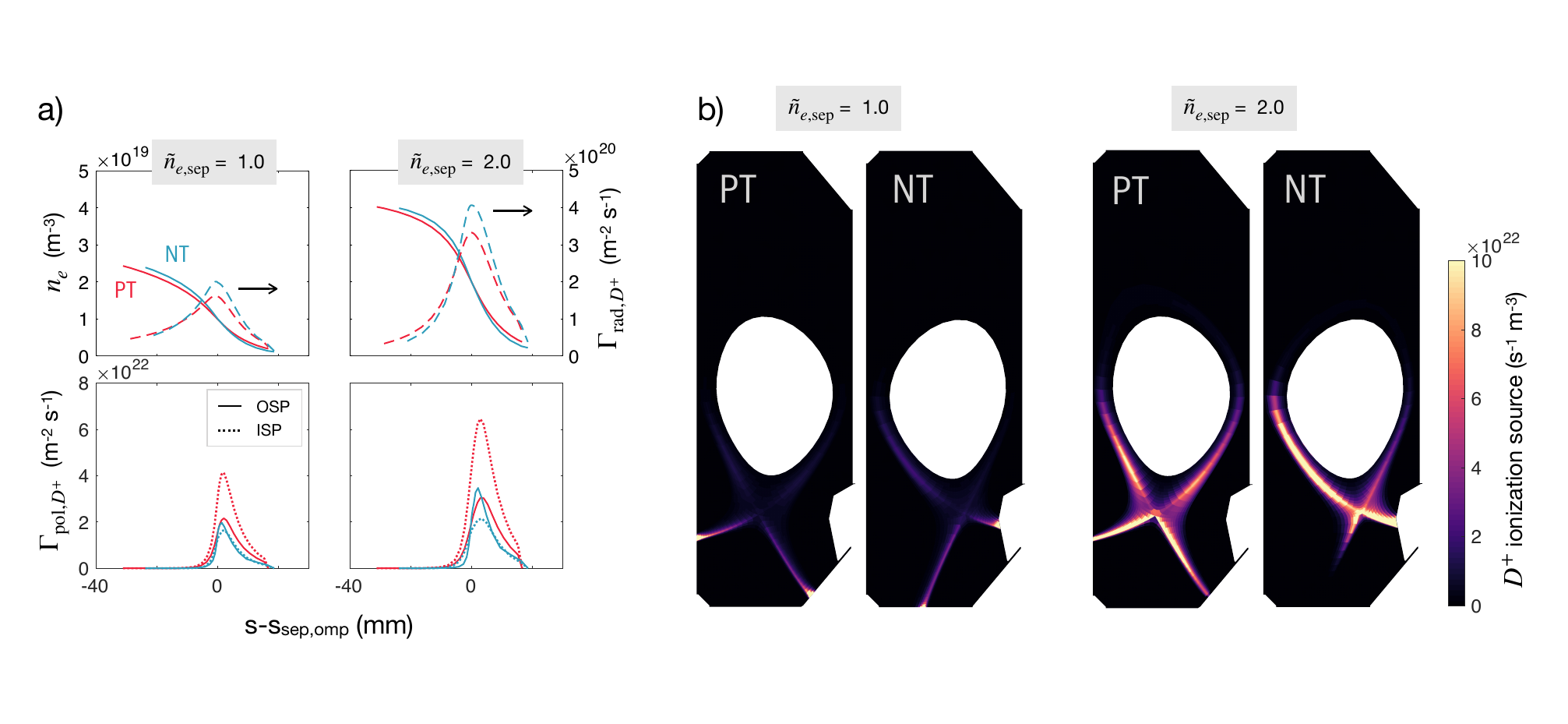}
\caption{a) Radial profiles of electron density (solid) and cross-field flux (dashed) at the OMP (top row) and poloidal fluxes at the inner (dotted) and outer (solid) strike points (bottom row) for PT (red) and NT (blue) configurations at two upstream densities $\tilde n_{e, \mathrm{sep}}=n_{e, \mathrm{sep}}/(1\times 10^{19}\, \mathrm{m^{-3}})$. b) Ionization $D^+$ source distribution in PT and NT in attached and partially-detached conditions.} 
\label{fig:NTvsPT} 
\end{figure*}

After comparing SOLPS-ITER reference simulations and the corresponding experimental data, in this section, we investigate the differences between PT and NT scenarios by comparing the two reference simulations at $n_{e, \mathrm{sep}}= 1.0 \times 10^{19} \, \mathrm{m^{-3}}$ (attached conditions) and two simulations for the same equilibria at higher upstream density $n_{e, \mathrm{sep}}= 2.0 \times 10^{19} \, \mathrm{m^{-3}}$. \\
This analysis focuses in particular on the comparison of radial and poloidal particle fluxes and ion sources from plasma-neutral interactions, which are the main actors in determining the particle balance. Figure~\ref{fig:NTvsPT}.a compares PT (red) and NT (blue) radial profiles of density (solid) and cross-field flux (dashed) at the OMP in the top row and the ISP (dotted) and OSP (solid) poloidal particle flux profiles in the bottom row. The left column refers to the low-density simulation while the right column to the high-density simulation. It is first noticeable that the upstream radial density profiles are not equal in PT and NT, although $D_{n}^{AN}$ is the same for both configurations. This is true both at low and high densities. In particular, the two density profiles intersect at $s=s_\mathrm{sep}$ because of the feedback scheme used in the simulation setup but the NT density is higher than PT in the core region ($s-s_\mathrm{sep}<0$) and lower in the SOL ($s-s_\mathrm{sep}>0$). This means that across the separatrix:
\begin{equation}
\left|\partial n_e / \partial r \right|_{NT} > \left|\partial n_e / \partial r \right|_{PT}
\end{equation}
In simulations without drifts and a single ion species $n_a \simeq n_{D^+} = n_e$, the cross-field flux (dashed line in figure~\ref{fig:NTvsPT}.a) is given by
\begin{equation}
\Gamma_{\mathrm{rad}, D^+} = -D_{n}^{AN} \frac{\partial n_e}{h_r \partial r}
\end{equation}
where $h_r$ is the metric coefficient in the radial direction. At fixed $D_{n}^{AN}$, the radial flux is proportional to the modulus of the radial density gradient. Setting the cross-field diffusion coefficients in mean-field edge codes like SOLPS-ITER, does not mean fixing the magnitude of the cross-field fluxes. Indeed, the radial density gradient can be different in simulations with the same $D_{n}^{AN}$, so that different $\Gamma_\mathrm{rad,n}$ may arise.  \\
The bottom part of figure~\ref{fig:NTvsPT}.a shows the profiles of the poloidal particle flux at the target (ISP with dotted lines and OSP with solid lines). Comparing PT and NT OSP profiles, simulations predict a narrower flux profile for the NT case compared to the PT one for both analysed densities. This is compatible with the observed narrower SOL width in NT compared to PT~\cite{Faitsch2018}. \\
Figure~\ref{fig:NTvsPT}.b shows the $D^+$ ion source distributions due to plasma-neutral interaction for PT and NT scenarios at the two densities of interest. At steady-state, particle fluxes and sources are related by:
\begin{equation}
\nabla \cdot \vec \Gamma_{D^+} = \sum_\mathrm{sources,sinks} S_{n, D^+}
\end{equation}
Different fluxes can thus arise if the plasma sources are distributed differently in the poloidal plane. The $D^+$ ionization source in the PT case is symmetric around the X-point while for the NT case, $D^+$ production is strongly asymmetric in favour of the high field side (HFS). Moreover, comparing the low and high-density simulations, we can observe the characteristics of an attached plasma, with the peak of the ionization source just in front of the target, in the first case while the latter moves more upstream at higher densities. The only fully detached target, according to the results of the simulations at  $n_{e, \mathrm{sep}}= 2.0 \times 10^{19} \, \mathrm{m^{-3}}$, is the ISP in the NT scenario. Comparing the high-density case for PT and NT, we can observe how the ionization front of the outer divertor leg is starting to move upstream in the former case while in the latter it is still strongly attached to the outer target. This is consistent with experimental observations.

\subsection{Density ramp modelling}
\label{subsec:densityRamp}
\begin{figure*}
\centering
\includegraphics[width=0.9\textwidth]{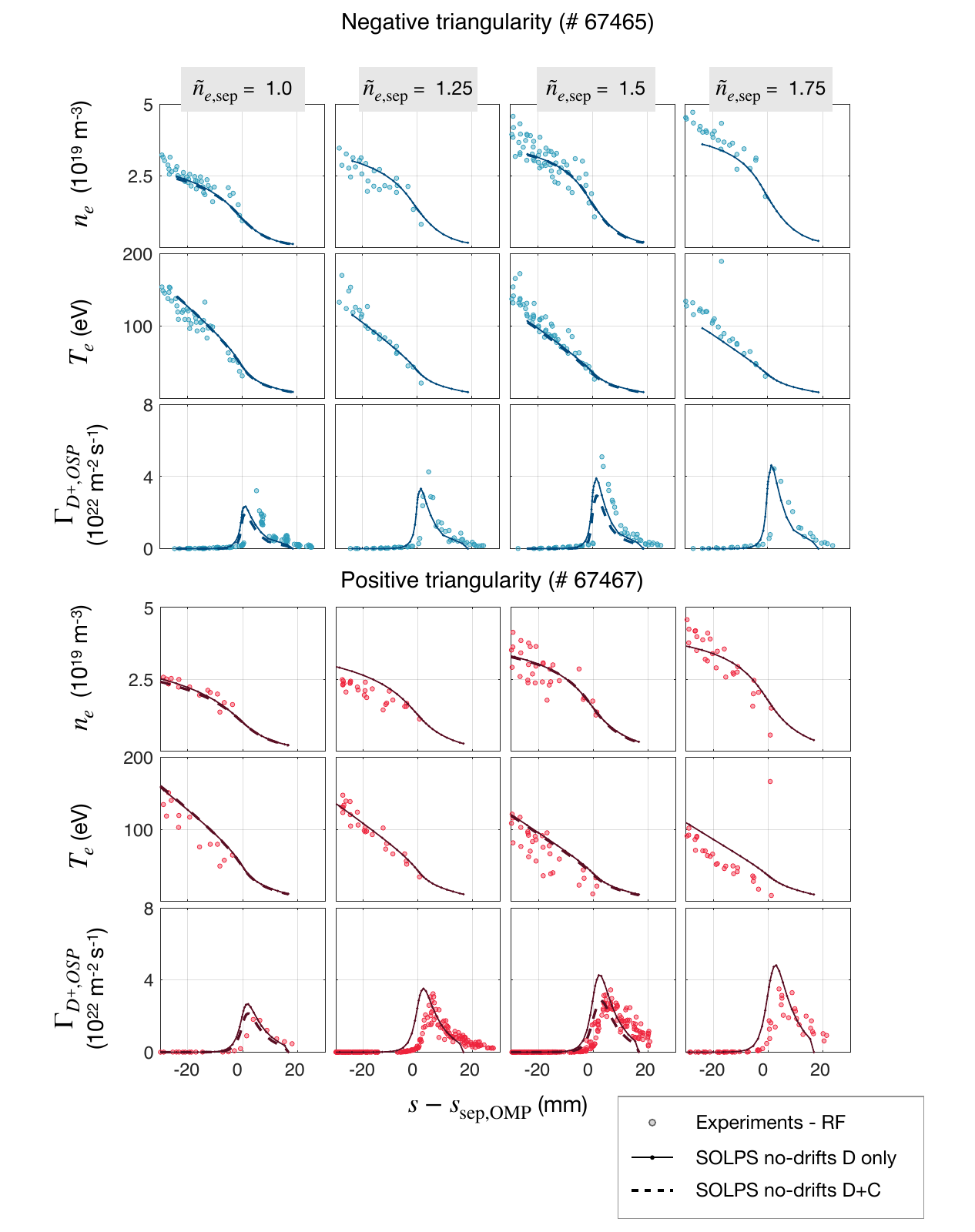}
\caption{Outer midplane density, temperature and outer strike-point ion saturation current profiles for different upstream densities. Profiles are plotted as a function of the upstream radial density from the separatrix. The three top panels (blue) represent the NT scenario and the bottom ones (red) the PT scenario. Upstream density is increasing from left to right according to $\tilde n_{e, \mathrm{sep}} = n_{e, \mathrm{sep}}/(1\times 10^{19} \, \mathrm{m^{-3}})$. Experimental data from TS (OMP profiles) and LPs (OSP profiles) are shown with dots, SOLPS-ITER profiles w/o C impurities (D-only) are shown with solid lines and, for $\tilde n_{e, \mathrm{sep}}=1.0$ and $\tilde n_{e, \mathrm{sep}}=1.5$, SOLPS-ITER profiles including C impurities (D+C) are shown with dashed lines. } 
\label{fig:PTramp} 
\end{figure*}
After the analysis and the validation of the reference attached scenario, this section presents the results of systematic density scan simulations in both PT and NT scenarios. These simulations aim to investigate the trends observed during density ramp experiments. Starting from the reference setup for NT and PT configurations described in section~\ref{subsec:RefSim}, we used the same input setup to scan the value of $n_{e,\mathrm{sep}}$ in the range $0.75 - 3.5 \times 10^{19} \, \mathrm{m^{-3}}$. All the other parameters, including anomalous transport, input power and boundary conditions, were kept fixed. As before, we isolated the effect of changing a single simulation parameter at a time. \\
Since we expect C impurities to play a role in the development of detachment, we included their effects for some of the considered upstream densities. C is produced via physical and chemical sputtering. We used energy and angle-dependent sputtering yields calculated according to the Roth-Bohdansky formula~\cite{Bohdansky1980} and a constant sputtering yield equal to $3.5 \%$, as done in previous TCV simulations~\cite{Wensing2019, Wensing2021}, respectively. \\ 
When C impurities are included, the maximum $n_{e,\mathrm{sep}}$ that allows to sustain the plasma for the reference input power is different in PT and NT:
\begin{equation*}
n_{e,\mathrm{sep, NT}}^\mathrm{max} = 2.25  \times 10^{19} \, \mathrm{m^{-3}} < n_{e,\mathrm{sep, PT}}^\mathrm{max} = 2.5  \times 10^{19} \, \mathrm{m^{-3}}
\end{equation*}
For this reason the maximum $n_{e,\mathrm{sep}}$ is different for PT and NT in figure~\ref{fig:Ttrends}.  \\

\subsubsection{OMP and OSP profiles}
The OMP density and temperature profiles and the OSP poloidal particle flux resulting from the density scan are shown in figure~\ref{fig:PTramp} and compared to TS and LP data, respectively. Upstream density is increasing going from left to right in the figure. The three top rows are for the NT scenario, while PT is on the bottom. In the panels for upstream densities $\tilde n_{e, \mathrm{sep}} = 1.0$ and $\tilde n_{e, \mathrm{sep}} = 1.5$ we also plotted the results of simulations including C impurities (dashed lines). The presence of C does not affect upstream profiles because of the feedback puffing scheme, however, to obtain the same $n_{e,\mathrm{sep}}$ in simulations with C, a lower gas puff is needed. A general good agreement between simulations and experiments is observed for the upstream profiles at increasing densities. For all densities, the trend observed section~\ref{subsec:RefSim} is mantained, i.e. SOLPS overestimates the upstream $T_e$ in PT employing the same $\kappa^{AN}_{e}$ and $\kappa^{AN}_{i}$ used for NT.\\
At the outer target, the effect of carbon impurities is to reduce the total ion flux by a combined reduction of $T_e$ and $n_e$, as can be observed from figure~\ref{fig:Ttrends}. As far as the agreement with experimental LPs data is concerned, simulations including C reproduce the trend of the OSP ion flux at increasing density in PT. In NT, the agreement with the experimental ion flux at the target increases with higher density although, when C is included, simulations underestimate the experimental data.

\subsubsection{Electron density and temperature at the strike points}
\begin{figure}
\centering
\includegraphics[width=0.5\textwidth]{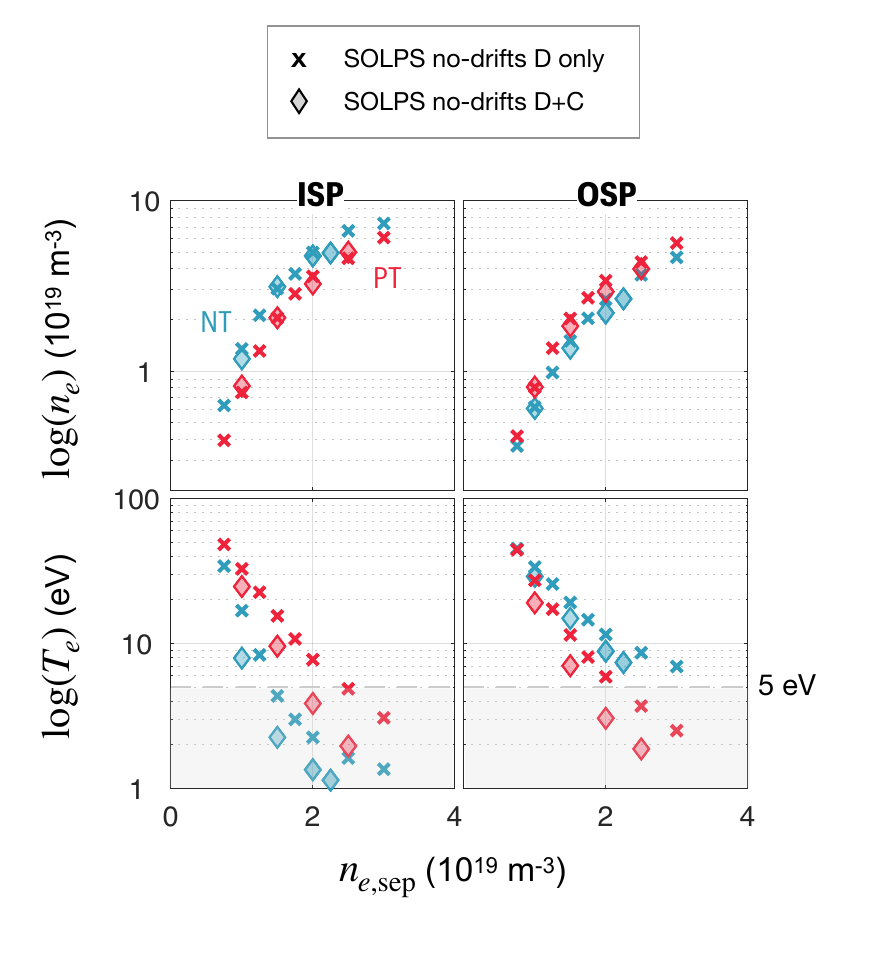}
\caption{Maximum electron density and temperature at the inner (left) and outer (right) strike points as a function of upstream electron density $n_{e, \mathrm{sep}}$. In the $T_e$ plots (bottom) a dashed line indicates $T_e=5$ eV, usually considered as a detachment threshold.} 
\label{fig:Ttrends} 
\end{figure}
Figure~\ref{fig:Ttrends} plots the maximum of density and temperature profiles in front of the inner (left) and outer (right) targets as a function of the upstream density. Again, red markers indicate PT and blue ones NT. Crosses are used for the results of simulations without carbon (D-only) while diamonds represent simulations including C (D+C). \\
From the lower right box in figure~\ref{fig:Ttrends}, we observe that simulations predict that the OSP electron temperature during a density ramp remains higher in NT compared to PT even assuming constant $D_n^{AN}$.  Moreover, even with the temperature reduction associated with C impurities, the maximum $T_e$ at the OSP in NT never falls below 5 eV (dashed grey line), which is often considered a first indication of detachment. In PT, on the contrary, the maximum $T_e$ is below 5 eV already for upstream densities $n_{e,\mathrm{sep}}\simeq 1.5 \times 10^{19} \, \mathrm{m^{-3}}$. This is consistent with what was observed experimentally~\cite{Fevrier2023}.

\subsubsection{Divertor neutral pressure}
\begin{figure}
\centering
\includegraphics[width=0.9\textwidth]{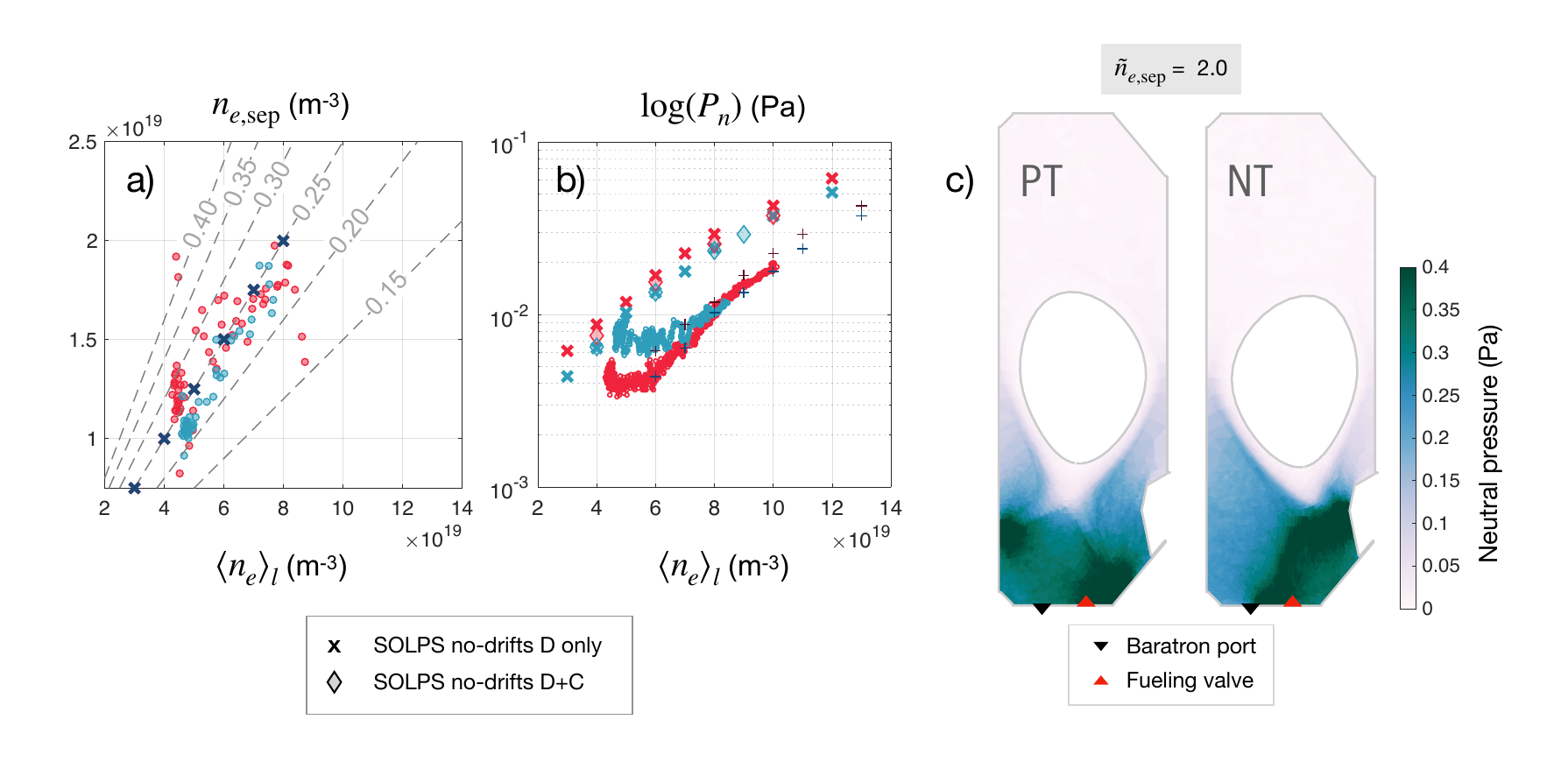}
\caption{a) Linear fitting of the upstream density $n_{e, \mathrm{sep}}$ as a function of the line-averaged density $\langle n_e \rangle_l$. To plot SOLPS-ITER results as a function of $\langle n_e \rangle_l$ the relationship $\langle n_e \rangle_l= n_{e, \mathrm{sep}}/0.3$ is used. b) Comparison of experimental divertor neutral pressure (dots), measured by TCV divertor baratron gauge, and synthetic pressure estimated from SOLPS-ITER results (crosses and diamonds, for D-only and D+C simulations respectively). + markers indicate data from SOLPS D-only simulations shifted by $\langle \tilde n_e \rangle_l = \langle n_e \rangle_l + 3 \times 10^{19} \, \mathrm{m^{-3}}$. b) Deuterium neutral pressure $p_{D_2}+p_D$ (contributions from molecular and atomic $D$) estimated by SOLPS-ITER simulations for $\tilde n_{e, \mathrm{sep}}=2.0$. } 
\label{fig:Pdiv} 
\end{figure}

Finally, we analyse the divertor neutral pressure. In figure~\ref{fig:Pdiv}.b, we show the experimental $P_{n, \mathrm{div}}$ as a function of the line averaged density $\langle n_e \rangle_l$, i.e. electron density averaged along a vertical line of sight of the Far InfraRed interferometer (FIR). In the experiments, $\langle n_e \rangle_l$ is the control parameter for the feedback loop that controls puffing during the core density ramp. From the point of view of power exhaust, however, a more relevant parameter than $\langle n_e \rangle_l$ is the upstream density $n_{e,\mathrm{sep}}$. Looking at the experimental $\langle n_e \rangle_l$ and $n_{e,\mathrm{sep}}$, we found an approximately linear relation between these two quantities, as shown in figure~\ref{fig:Pdiv}.a. To plot SOLPS results as a function of $\langle n_e \rangle_l$, we thus used the relationship $\langle n_e \rangle_l^\mathrm{SOLPS}=n_{e,\mathrm{sep}}^\mathrm{SOLPS}/0.25$ for both NT and PT, consistent with previous TCV studies~\cite{Fvrier2021, Wensing2021}. \\
To compare the baratron measurements with SOLPS results, we need to make assumptions on the neutral transport from the TCV chamber to the baratron gauge, located outside the tokamak magnetic field~\cite{Wensing2019}. Differently from what was proposed in~\cite{Wensing2019}, we relaxed the hypothesis of isotropic thermal fluxes entering the baratron channel from the divertor side, taking instead the kinetic fluxes estimated by EIRENE. This reduces the discrepancies between simulations and the baratron data by almost a factor of 2. SOLPS-ITER simulation reproduces the experimental trend but the results are shifted to lower densities (figure~\ref{fig:Pdiv}.b). Qualitatively, this is compatible with a time delay between the actual pressure increase in the vessel and the baratron measures. More quantitative analysis is planned for future studies. \\
Because of the different divertor geometries, comparing the baratron measures for PT and NT gives information that is difficult to interpret. Nonetheless, it is interesting to notice how neutrals distribute differently in the two configurations (figure~\ref{fig:Pdiv}.c). As for the ionization $D^+$ source, discussed in section~\ref{subsec:diffPTNT}, the neutral pressure distribution is quite different between PT and NT configurations. For the PT scenario, the neutral pressure peaks at the strike points, it remains high in the private flux region (PFR) and then decreases symmetrically going towards the midplane across the X-point. For the NT configuration, instead, the neutral pressure peaks along the whole PFR, all the way towards below the X-point. In addition, as for the ion ionization source, it is strongly asymmetric between HFS and LFS regions. 

\section{Conclusions}
\label{sec:conclusions}
This paper presented the comparison and analysis of two TCV discharges with opposite triangularity through SOLPS-ITER modelling. The main goal of the work was to provide insights into the physical reasons behind the results of recent detachment experiments done in TCV, which found that detaching the outer divertor leg in NT is harder than in PT~\cite{Fevrier2023}. The analysed discharges are two LSN Ohmically heated L-mode with opposite lower and upper triangularity. Reversing both upper and lower triangularities produced a different divertor geometry for the two configurations. \\
The analysis with SOLPS-ITER was done at constant input parameters for both PT and NT. In particular, input power, boundary conditions and anomalous diffusivities for particles $D_n^{AN}$ and energy $\kappa_{e/i}^{AN}$ were kept fixed. The latter assumption is meant to fix one of the degrees of freedom of mean-field edge transport codes - like SOLPS-ITER - which do not consistently describe anomalous transport and avoid ad-hoc assumptions on the magnitude of these parameters for the different triangularities. The results of the simulations, nonetheless, recovered many interesting features observed in experiments presented in~\cite{Fevrier2023}. At the same upstream plasma density $n_{e, \mathrm{sep}}$, we found a higher core density for the NT scenario due to a lower cross-field particle flux inside the separatrix. Despite the fixed anomalous diffusivity, a steeper gradient of the density profile results for the NT case, indeed imposing the same $D_n^{AN}$ in two different mean-field simulations does not imply recovering the same magnitude of cross-file transport. \\
In scenarios with different divertor geometries, the recycling of neutral particles at the wall, their transport within the plasma and the consequent distribution of the plasma ionization source play the major role in causing the differences observed between the PT and NT scenarios. To disentangle the role of neutral recycling and transport and possible upstream differences in cross-field transport, future studies should be performed on configurations that only differ in the upper triangularity while having the same divertor geometry as already explored experimentally~\cite{Fevrier2023}. \\
Particularly relevant are the differences observed in terms of peak electron temperature at the OSP. This parameter was shown to be a robust quantity under the simplifying assumptions used in this work and the trends recovered in the simulations well reproduce the experimental observations~\cite{Fevrier2023}. Simulations found that the outer target is more difficult to detach in NT compared to PT. For the same upstream density, indeed, a hotter OSP is found in NT than in PT. Moreover, at increasing upstream density, the maximum OSP $T_e$ for the NT geometry never fell below the detachment reference temperature of 5 eV, consistent with what was quantitatively observed in TCV experiments~\cite{Fevrier2023}.

\section*{Acknowledgements}
This work has been carried out within the framework of the EUROfusion Consortium, partially funded by the European Union via the Euratom Research and Training Programme (Grant Agreement No 101052200 — EUROfusion). The Swiss contribution to this work has been funded by the Swiss State Secretariat for Education, Research and Innovation (SERI). Views and opinions expressed are however those of the author(s) only and do not necessarily reflect those of the European Union, the European Commission or SERI. Neither the European Union nor the European Commission nor SERI can be held responsible for them.
The results are obtained with the help of the EIRENE package (see www.eirene.de) including the related code, data and tools~\cite{Reiter2005}.

\section*{Reference}
\bibliographystyle{iopart-num}
\bibliography{Bibliography.bib}  

\end{document}